# WHAT DO WE TEACH TO ENGINEERING STUDENTS: EMBEDDED ETHICS, MORALITY, AND POLITICS

Avigail Ferdman, Department of Humanities and Arts, Technion Israel Institute of Technology

Emanuele Ratti[1], Department of Philosophy, University of Bristol,

Authors have contributed equally and are listed in alphabetical order.

**Abstract.** In the past few years, calls for integrating ethics modules in engineering curricula have multiplied. Despite this positive trend, a number of issues with these 'embedded' programs remains. First, learning goals are underspecified. A second limitation is the conflation of different dimensions under the same banner, in particular confusion between ethics curricula geared towards addressing the ethics of individual conduct and curricula geared towards addressing ethics at the societal level. In this article, we propose a tripartite framework to overcome these difficulties. Our framework analytically decomposes an ethics module into three dimensions. First, there is the ethical dimension, which pertains to the learning goals. Second, there is the moral dimension, which addresses the moral relevance of engineers' conduct. Finally, there is the political dimension, which scales up issues of moral relevance at the civic level. All in all, our framework has two advantages. First, it provides analytic clarity, i.e. it enables course instructors to locate ethical dilemmas in either the moral or political realm and to make use of the tools and resources from moral and/or political philosophy. Second, it depicts a comprehensive ethical training, which enables students to both reason about moral issues in the abstract, and to socially contextualize potential solutions.

1. INTRODUCTION

Beth, a computer scientist, and Tom, a philosopher, are designing an ethics module for their 'Robotics and Autonomous Control' course, taught at a school of engineering. Their aim is to instill in their students an 'ethically responsible mindset'. Ethics modules are emerging as a tool in teaching ethics in engineering schools. In particular, embedded ethics is a popular pedagogical strategy that is used to teach ethics not as stand-alone modules, but in direct connections to technical modules, in order to associate more tightly the ethical and the technical. In particular, embedded ethics pedagogy aims to train students to (Grosz et al. 2019):

1. identify potential harmful consequences and other ethical problems early in the design process;
2. take steps to eliminate or mitigate them;
3. negotiate among competing values;

---

[1] mnl.ratti@gmail.com

4. reason about those problems and potential solutions to them, using concepts and principles from moral philosophy.

One of the pedagogical challenges of instilling an ethically responsible mindset among engineering students is getting the students to recognize that they have an ethical responsibility in the first place (Kopec et al. 2023). Embedded ethics' strategy to overcome this challenge is to habituate students to thinking ethically as they develop algorithms and build systems, by using recurrent exposure to ethical problems throughout the core curriculum of computer science (Grosz et al. 2019). Despite the increasing number of modules available, some issues have emerged. The first is that it is not clear what the learning goals and outcomes of such modules are. Is it about introducing some (ethical) concepts and exposing students to them? Is it about reflecting on professional duties? Or is it about cultivating some skills related to civic and moral standards, on a par with technical skills? Standards on how to design them and thorough reflection on their nature in terms of learning goals are lacking. A second issue is that individual and collective dimensions of engineering practice are often not clearly distinguished (Franssen, Lokhorst, and van de Poel 2018), potentially leaving questions of responsibility and accountability unresolved, and obscuring the moral duty to come up with solutions at the societal level. In the case of Beth and Tom, addressing these issues would amount to answering questions such as: how should they go about designing and teaching the module? What are exactly the goals, and which strategy should they put in place to achieve them?

In this article, we propose that to successfully accomplish embedded ethics aims, Beth and Tom ought to design the module with the following emphases in mind: Instilling ethical and moral reasoning through *habituation*; understanding and applying moral concepts; and anticipating the moral consequences in their societal context. These emphases, we argue, should be explicitly formulated as three distinctive parts of ethics modules: the ethical dimension (Section 2), the moral dimension (Section 3), and the political dimension (Section 4). The ethical dimension will deal explicitly with learning goals, which is to make ethical reasoning 'characteristic' of engineering practice. The moral dimension refers to the content of what is made 'characteristic', and in particular to the moral content. While the ethical and the moral dimensions have (implicitly) been at the core of embedded ethics content (see for example (Grosz et al. 2019; Pasricha 2023), the political dimension has received less attention. The political dimension concerns what we all together as members of a community owe other members of the community, when we act in and on behalf of the artificial collective person which is the state (Dworkin 2011), including the legitimacy of exerting power and agreement over legitimate disagreements (Miller 1998; Rossi and Sleat 2014). This dimension is important when teaching applied ethics to engineers, because as future practitioners, they will be designing algorithms and AI that have the power to shape whole social systems (Vallor 2021). Additionally, many of the issues we face (e.g. self-driving cars' behaviour) are collective decisions that a society makes together, mediated through institutions and policies, rather than individual moral decisions we make in light of what we each have reason to value (Himmelreich 2020). The political dimension helps to make these aspects of technology explicit, and enables us to utilize tools from political philosophy and political theory for addressing them.

In what follows, we will describe this tripartite framework, and we will show how one can design an embedded ethics module on the quite popular topic of autonomous vehicles.

In addition to illustrating the framework through the example of autonomous vehicles, we will also argue for some of the strengths of the framework. In particular, our approach provides analytic clarity: it enables course instructors to locate ethical dilemmas in either the moral or political realm, and to make use of the tools and resources that moral or political philosophy provide accordingly. A second advantage is that it provides a comprehensive ethical training, which enables students to both reason about moral issues in the abstract, and to socially contextualize potential solutions.

Before describing our tripartite framework, it is important to provide a short list of the challenges of autonomous vehicles. These will turn out to be useful to provide concrete examples on how the different dimensions of a hypothetical ethics module can be concretely designed.

There are multiple ethical challenges and dilemmas associated with autonomous vehicles. These challenges are manifest in the design of the behaviour of the single autonomous vehicle unit (e.g. designing safe and fair algorithms, balancing accuracy and justice in data training schemes). The challenges also manifest in sociotechnical issues such as decision-making processes about the behaviour of the autonomous vehicle as a system; transportation and road infrastructure and the political economy of route optimization (see for example (Ferdman 2020). The three-dimensional framework (ethics, morality, politics) provides an analytic framework for Beth and Tom, within which to design their ethics module, thus addressing both the ethical dilemmas implied by the autonomous vehicle technology, and the social responsibility of students to mitigate potential ethical problems associated with this technology.

## 2. The Ethical Dimension

The first dimension of teaching ethics in the context of engineering is about teaching goals. Often, these are loosely based on taxonomies conceived along the basis of Bloom's taxonomy of educational objectives (Bloom 1956) or any of its subsequent developments. Bloom's and similar taxonomies have different learning domains, which include different levels of learning. The cognitive domain is the most relevant here, and it includes learning goals such as remembering, understanding, applying, analyzing, evaluating, creating, etc. Usually, these broad goals have to be filled with content, e.g. 'understanding possible worlds semantics'. In the case of teaching ethics to engineers, specifying the learning goals is important to counteract the popular view that ethics curricula in technical disciplines are just taught for compliance reasons. For instance, in the US researchers receiving NIH, NSF, or USDA fundings have to undergo ethical training; and this may sometimes send the message "that education in ethical and responsible research practice is merely a hurdle to be cleared" (Chen 2021, 229). We could not find any in-depth reflection on learning goals for ethics courses, other than just vague reference to the taxonomies of learning goals mentioned above.

What does this lacuna have to do with the fact that we call this first dimension 'ethical'? We think about the 'ethical' in the sense of *ethos*, which can be understood as 'characteristic' (Russell 2014). The learning goal of teaching ethics to engineers will then be to facilitate a process whereby one becomes accustomed to do something or where doing something becomes 'characteristic' (Russell 2014), where this 'something' is related to the moral or political content (see Section 3 and 4) of the course itself. In other words, the goal is to make students familiar with certain activities in a way that these activities become *ethoi.* More precisely, the goal of teaching applied ethics is to suggest a way in which some activities can become 'ethoi', and to inspire students in such a way that they will be actively pursuing those activities to get to the level of becoming 'habits'. Such learning goals are obviously hard, but not implausible. But these 'habits' should not be conceived as passive absorption. As the reader has probably guessed, the way we describe the ethical dimension comes out of the virtue ethics tradition. This tradition has shown very forcefully that becoming accustomed to do something must also come with the ability to recognize and deliberate about the reasons for acting in a way rather than another (Annas 2011). Our proposal takes inspiration from virtue ethics in order to talk about learning goals, and it commits to virtue ethics only from this instrumental point of view. We do not want to subsume engineering ethics under virtue ethics, and for this reason we do not need to defend virtue ethics against typical objections, such as situationist critiques (Sreenivasan 2013).

*2.1 Learning Goals as Cultivating Virtues*

In order to understand the learning goals we are talking about, we need to introduce the language of virtues.

The very meaning of virtue is a controversial topic that cannot be solved in a short article, and this why here we will use a working definition. A virtue, in its original formulation (Russell 2014) is an excellence. This is a characteristic in virtue of which something or someone is deemed good. One can talk about excellences by referring to animals, objects, etc. There are excellences that are typically human's. For instance, one can think about being an excellent guitar player or an excellent skier. When it comes to technical disciplines such as computer science, excellences can be good coding or good programming. These kinds of excellences are called *skills*. In philosophy, intellectual, moral, and civic skills are called 'virtues', and can include curiosity, honesty, and fairness respectively. The difference between skills and virtues has been a topic of controversy since Aristotle formulated it, but the main idea is that skills generate products whose goodness is to be established by means of their internal characteristics, while products coming out of virtues are good "not simply if they are in a certain state, but if the one who does them is also in a certain state" (EN II.4). But what virtues and skills have in common is far more important for this article than how they differ.

Characteristics that are shared by these different excellences (be they skills, or virtues) is that they are stable character traits, or long-lasting ways at being good. Stability is important because one cannot be considered an excellent skier if he/she skies well once – the trait must be 'characteristic' of that person. Most important, the way in which these can become characteristic and stable is through an active process of learning and habituation. Courses in disciplines such as computer science have skills as learning goals; these can include proficiency in programming languages, design of database systems, coding, data cleaning, etc. These skills are taught with a mix of theoretical knowledge and lots of practical exercise. *Mutatis mutandis*, we should think about ethics courses in technical curricula in the same way: their learning goals should be facilitating the cultivation of certain intellectual, moral, and civic excellences (i.e., the virtues), at least ideally. Using the language and concepts from virtue ethics then is a powerful way to think about learning goals. However, a caveat should be added. In the 21[st] century, the word 'virtue' "can conjure images of Victorian patriarchy or dusty children's books filled with earnest morality tales"(Ratti and Stapleford 2021, p. 1).

Because we want to avoid this, while the ethical dimension is conceptualized by instructors through the language and concept of virtues, the way it is described to students can avoid using words like 'virtues', and go instead for words like 'skills', 'excellences', which are more neutral and less controversial.

*2.2 Advantages of Conceptualizing Learning Goals as Virtues*

Conceptualizing learning goals through the language of the virtue tradition has several advantages.

First, focusing on virtues as learning goals can specify more precisely the general goals that are usually ascribed to ethics modules in engineering curricula. For instance, Johnson (2017) says that one goal is to make students "aware of what will be expected of them in their work as engineers" (p. 61), which can be expressed as cultivating certain civic virtues related to how practitioners should behave with respect to their professional communities and to the wider public. Another goal is to "sensitize students to ethical issues" (p. 61), which can refer to the cultivation of a moral sensibility or attention that facilitates students' perception of their own work from a moral standpoint. Another goal is to improve "students' overall ethical decision making and judgment" (p. 61) which can be expressed by referring to moral virtues, as well as intellectual virtues, which are necessary for good reasoning. Cultivating moral, intellectual, and civic virtues allows students to treat ethics not as an inconvenience they have to comply with, but rather as part and parcel of their own (future) professions.

Next, formulating learning goals as virtues will also make more explicit the ways in which such goals can be (in principle) achieved. Virtues are cultivated exactly in the same way that technical skills are cultivated: by means of exercises. For instance, in (2019) Kohen et al describe four crucial components of 'heroic' people, which can be used to specify learning goals representing the cultivation of any virtue (be they moral, intellectual, or civic). In the context of, for instance, teaching AI ethics, students must (1) develop skills (i.e., virtues) or perceptions that are useful to address ethical and civic concerns about AI tools; these skills/virtues may be cultivated by (2) regularly taking action that develop those skills/virtues, such as (3) in the continuous imagination of situations in which these skills can be exercised. These should lead (4) to an expansive sense of morality and, we add, understanding of how AI tools shape the whole society. These four aspects emphasize the need for an active-learning (Bezuidenhout and Ratti 2021) and activity-based style of teaching ethics to engineers. As philosophers, we are tempted to teach what we know through theoretical means; however, we should just take what we know as a background and then engage the students with exercises that habituate them to certain activities and stimulate the cultivation of ethical, intellectual, and civic excellences. In other words, as instructors we should facilitate a sort of ethical, intellectual, and civic maieutic process, rather than *ex cathedra* show off our philosophical skills. Whether a course can achieve, at least ideally, certain learning goals formulated in terms of excellences, should be evaluated by taking into account the activities planned to cultivate those excellences, rather than the mere content of the course itself.

Another advantage of focusing on virtues to specify tangible learning goals is that the cultivation of moral, intellectual, and civic excellences can address a limitation of current professional ethical codes, which are often used to teach 'ethics' in professional, academic, and institutional contexts. As Harris notes (2008), professional codes of ethics have a substantial component of 'preventive ethics', an inclination towards prohibitions, and even provisions that, even if they are not stated negatively, they "nevertheless have an essentially negative force" (p. 154). Preventive ethics is usually formulated in terms of rules, which can be problematic for a number of reasons. First, the emphasis on rules suggests that all you need for being 'ethical' is to know the rules themselves, which can be mindlessly applied (Ratti and Graves 2021). This 'compliance' paradigm of rule-based ethics forces a legalistic interpretation of rules, which in turn promotes loop-hole reasoning and a perception of ethics as externally imposed (Kelly 2018). The second problem is that rules cannot possibly cover any possible situation, and different contexts may require "discretion, judgment, and background knowledge in meeting some professional obligations" (Harris 2008, p. 155). Only an approach based on virtues as learning goals can equip students and future professionals with the moral sensibilities, intellectual skills, and civic understanding to address a wide range of contexts[2].

Moreover, a focus on cultivating virtues fits nicely with dispositional conceptions of the trustworthiness of professions. Such accounts (e.g. Jones 2012; Kelly 2018) emphasize that a member of a professional community is trustworthy when "the dependence and vulnerability of the trustee counts as a compelling reason for them to responsibly care for the entrusted interest" (Kelly 2018, p. 45). Recognizing that vulnerability is a compelling reason, and being moved by it, implies that some professional ethics are, at their core, *ethics of care*, and that character traits are fundamental in this endeavor. Courses in engineering ethics that explicitly promote the cultivation of character traits necessary for recognizing and being moved by 'vulnerability' are fulfilling an important role for engineering communities: give students a taste of what it means to be a trustworthy member of a professional community. In the case of the ethics module for the course on autonomous vehicles, Beth and Tom could habituate an 'ethics of care' by showing how vulnerable social groups might be disproportionately affected in accident scenarios, such as cases where AI training datasets are much less sensitive to darker skin tones (Buolamwini and Gebru 2018). Seeing as the autonomous vehicle predictive performance might be discriminatory against pedestrians with darker skin tones (Wilson, Hoffman, and Morgenstern 2019), the module could emphasize the responsibility of the students as future members of their professional community to recognize such vulnerabilities and mitigate them in the design process.

Finally, emphasizing virtues in learning goals speak directly in favor of 'embedded ethics' approaches (Bezuidenhout and Ratti 2021; Grosz et al. 2019; McLennan et al. 2020; 2022). This is because the goal of embedded ethics approaches is to promote the learning of ethics not simply as things that one have to be 'exposed' to in the form of theories, principles, etc, for mere compliance reasons; rather, ethics is considered as something that is exercised and learnt as if it were a skill belonging to the profession itself, on a par with other technical skills. In other words, at the heart of embedded ethics approaches is the intuition that ethics is something like a skill, which is exercised and cultivated much like other technical skills.

---

[2] We are not suggesting that we should ignore codes of conduct completely. Actually, we think that they are an excellent way to 'negotiate' the ethos of a community. What we want to avoid is the temptation of interpreting codes of conducts merely in terms of rules rather than, in general, as something documenting the ethos and general principles endorsed by a community of professionals.

Formulating learning goals in terms of virtues that professionals have to cultivate gives to this intuition a much more stable ground.

All in all, these are reasons for instructors to conceptualize learning goals by using a virtue mindset.

*2.3 Which virtues?*

What are the virtues that should be cultivated and how exactly these virtues should be understood is an open question.

We can draw from general professional ethics (Kelly 2018) and mention loyalty, beneficence, respect for autonomy, honesty, discretion, etc. However, the landscape of professional virtues is vast, including dozens of virtues in many different contexts, from military, leadership, business, AI, to medicine just to mention a few (Hagendorff 2022). Moreover, given the limited time and format of ethics modules, rather than aspiring to full-blown virtues (whose actual existence is sometimes questioned), we set as a goal the development of two 'perspectival sensibilities', which we define as *educated ways of seeing things from a certain perspective*. Let us unpack this.

Engineering students, and in particular computer science students, are taught to see their objects of design and inquiry in a purely technical way, and the exercises and puzzles that they have to solve address computational objects only from a technical point of view. This is, of course, for a very good reason: eventually, students must be able to design computational systems, and this requires the cultivation of technical skills. However, by embedding ethics modules in engineering curricula, it is possible to enlarge the range of perspectives through which technical objects like computational systems are considered. Perspectivity includes different dimensions, in the sense that an object is investigated in light of certain problems and questions, under certain assumptions, and with certain goals in mind. For instance, in the case of a technical perspective, the training of a machine learning system can be seen in light of problems of quality of data, computational power, whether the goal to achieve can be formulated as a classification problem, etc. By practicing and practicing at seeing and manipulating machine learning systems under the lens of this technical perspective, novices become data scientists or, at least, technically skilled data scientists. In ethics modules, we want students to see technical systems also from different perspectives. In particular, the learning goal is to develop a sensibility that allows them to see algorithmic systems from both a moral and a political point of view. And one way to do this, analogous to the cultivation of skills, is to practice in seeing algorithmic systems under the lens of both a moral and a political (or, as we will call it, civic) perspective. For example, in the context of training an autonomous vehicle behaviour algorithm, this same object can be seen from the point of view of concerns related to how it impacts the safety and autonomy of passengers and pedestrians; how it can promote structural inequalities; how it impacts the environment. The goal of ethics modules should be to stimulate students to see the same technical objects also from a moral and a political point of view. This 'seeing' is the sensibility we talked about.

As mentioned above, it is unlikely that with the limited space of ethics modules students will master the moral and the political in the same way they master the technical - in other words, the cultivation of full-blown virtues, as analogous to full-blown skills, is improbable. However, we think that at least a perspectival sensibility - defined as an *educated way of seeing things from a certain perspective* - can be achieved. We call these perspectival sensibilities *moral attention* and *civic attention* which constitute, respectively the second (the moral) and third (the political) dimension. In Sections 3 and 4, we explain in detail the nature of these sensibilities.

## 3. THE MORAL DIMENSION

The previous section outlined the pedagogical goals of applied ethics courses. In particular, we have said that we should formulate learning goals from the point of view of the virtue tradition. From this angle, the goal is to cultivate certain abilities that allow us to master the typical topics of ethics modules as they emerge from technical practice. Given the difficulties of cultivating full-blown virtues in the limited space of a few modules, we have settled for the cultivation of a perspectival sensibility, which is an educated way of seeing things from a specific angle. In this section, we describe the content of the 'perspectival sensibility' that we call 'moral', and we shift the focus from the (moral) agent to the content, in this case moral content.

The moral dimension pertains to the moral norms that individuals ought to follow or that they think ought to follow. 'Moral norms' here refer to claims about right and wrong conduct. 'Right and wrong' is conceptualized as in relation to certain impacts that behaving in a way or another might have in shaping humans' lives or in shaping conceptions that people have of how they ought to live their own lives. Morality, as such, is a specifically interpersonal normative order (Setiya 2022), of "complete virtue … in relation to another" (NE 1129b26-28). For instance, assaulting individuals in your own neighborhood is a conduct that is morally relevant, because of the consequences for the lives of individuals who are victims of violent actions (i.e. they are harmed), but also because such a conduct can potentially shape other people's conceptions of how they ought to live their lives (i.e. living in fear), and as a consequence their freedom in pursuing what they see as desirable. We say that a conduct is morally relevant when it can be conceptualized in terms of the impacts briefly explained above. The moral attention or 'sensibility' that we refer to in Section 2 is then the ability to see how a certain conduct can be morally relevant. In an ethics module, the instructor will have to design activities or exercises that will stimulate students seeing the moral dimension in their own technical work

One way to get familiar with the moral dimension, is to exploit pre-existing resources that make available moral content, which can be used as proxies for identifying the moral relevance of engineering systems. Ideally, those resources should also provide strategies to investigate the moral content. Most important, these resources should be intuitive and of an easy grasp for those who do not have a strong philosophical background and do not have the time to engage in specialized debates. There are different traditions that fit this profile. For instance, one can use the list of values used in the literature of value-sensitive design (Friedman and Hendry 2019, 28) as a proxy to identify moral content. Another example – which is the one we develop here – is using principlism. We hasten to add that we use principlism only instrumentally and to motivate our pedagogy: it is a neat, precise, and straightforward way to talk about values, and to solve conflicts between values.

Principlism has a long-standing reputation, especially in the context of biomedical ethics (Beauchamp and Childress 2009), and responsible conduct of research (Shamoo and Resnik 2015). It has been a response to the several difficulties of doing bioethics by appealing to 'high moral theories' (e.g. which theory should we use? Which version? How are theories applicable concretely?). Unlike high-level theories, principlism is a normative framework based on mid-level moral norms (called *principles*); these "express general norms of the common morality (…) should function as general guidelines for the formulation of the more specific rules" (Beauchamp and Childress 2009, 12), and they have the advantage of being easier to interpret (Shamoo and Resnik 2015). It is worth noting that, among the applied ethics in the engineering context, AI ethics seems to have drawn inspiration for much of its content from principlism, especially biomedical ethics. But this move is not without its critics. While addressing these criticisms may seem beyond the scope of this article, doing this will make the case for why principlism is a good candidate for the moral dimension in teaching engineering ethics courses.

The first criticism argues that the principles of biomedical ethics are specific to the biomedical context, which is rather different from the AI context: hence, the import of certain principles in AI ethics is suspicious (Mittelstadt 2019). To answer this, consider the original motivations that Beauchamp and Childress had in developing this project in the biomedical context. They often refer to the idea of common morality, and they claim that this is the source of four universal and basic ethical principles: respect for autonomy, nonmaleficence, beneficence, and justice (2009). What this means is that, in their view, the four principles of biomedical ethics represent "the most general and basic norms of the common morality" (Beauchamp 2007, 7). This means that, while the biomedical context is different from the AI context and other engineering contexts, we can still use the same principles, as the basic concerns of common morality will apply across several contexts. In the specific case of AI, one can easily see how principlism can be applied regarding concerns about respect for autonomy (e.g. the way recommender systems operate), nonmaleficence (e.g. how AI systems can actually harm individuals in various ways), beneficence (e.g. whether AI systems can actually improve well-being), and justice (e.g. the misuse of risk assessment algorithms in the justice system). In other words, the same basic concerns of common morality will apply also to the context of various engineering courses. We can use the content of the principles as proxies for identifying the moral relevance of engineering systems. Using again the example of AI, one simple heuristics is to identify the main tasks that one has to do to design an algorithmic system, and then have group discussions on how the moral concerns represented by the principles can possibly apply. In this way, students get familiar with the task of connecting the moral concerns with the actual technical characteristics of engineering systems. Going back to the case of Beth and Tom, they could design their module to discuss the relevance of the four basic principles (respect for autonomy, nonmaleficence, beneficence, and justice) to a technical task such as designing a reward function for the autonomous vehicle's route optimization training. Most important, Beth and Tom can use different ways of conceptualizing the principles (which is something that principlism has emphasized considerably) as a way to guide discussion.

The second criticism states that the principled approach does not provide any guidance to practitioners on what to do, especially when conflicts arise. This can be addressed by considering that principlism in AI ethics is a rather impoverished version of traditional principlism. The objection that principles underdetermine moral deliberation and hence applicability is old news. This underdetermination challenge motivated the development of the so-called specification principlism (Beauchamp 2007). This is in direct opposition to two common strategies for 'using' principles – application (i.e. deductive subsumption) and 'situational intuition' (Richardson 2000). The former is a version of deductive subsumption of a case under the principle, but this is rather difficult to do, given the different (and sometimes conflicting) interpretations of the principles themselves, and the nature of the subsumption itself. The second is a sort of *phronesis*, but as such it leaves "the reasons for decision unarticulated" (Richardson 2000, p. 287). Both strategies have been used (implicitly) in AI ethics and, given the criticisms to the principled approach in this context, they have not been very successful. Moreover, both strategies address only the problem of bringing norms to bear in specific contexts, and leave out the problem of conflicts between norms. Specification principlism claims that, in order to fill the gap between general principles and concrete deliberation, one has to reduce the indeterminateness of general principles, in particular by narrowing down their scope. This is done, for instance, by being more precise about "where, when, why, how, by what means, to whom, or by whom the action is to be done or avoided" (Richardson 2000, 289). In an ethics module, one can use various strategies of specification to understand whether a general principle or value that is being discussed is really relevant to a specific technical task. It is important to emphasize that the specification should be carried out in a discursive way, by taking into account the technical nature of the tasks behind the design of engineering systems; this is yet another way to connect the moral dimension with the technical dimension. By narrowing down the principle to bear to the particular technical context, one can stimulate a discussion over ways in which the specified version of a principle can be promoted or obfuscated by certain technical moves. It is also important to add that one can also enlarge the set of principles without going towards a first round of specification. For instance, in value-sensitive design there are extensive lists of values that can be promoted in design, which can be easily formulated in terms of principles (Friedman, Kahn, and Borning 2015). Going back to autonomous vehicles, Beth and Tom's module could expose the students to trolley-like problems and foster a deeper appreciation of the different (potentially incompatible) moral conceptions that are inevitably built into the behavior of the AI. In alternative, one can also consider the different formulations of basic principles as they appear in typical AI ethics literature (Jobin et al 2019) and discuss how they relate to technical tasks. However, the pedagogical appeal of traditional principlism should not be overlooked: the four principles of biomedical ethics are so basic and intuitive that can be used as excellent 'hooks' for discussion by those who have not a philosophical training.

Relatedly, traditional principlism also provides resources to overcome the common phenomenon of principles/values conflicts, even though in a limited way. This is usually illustrated by striving for a wide reflective equilibrium (Richardson 2000), where the conflicts between abstract principles or values are defused by specifying those conflicting principles until they are in 'equilibrium' with each other (i.e., when they are coherent). For example, in algorithm-based sentencing, or algorithmic assessment of recidivism (Larson et al. 2016; Petersen 2021) there might be a conflict between accuracy (e.g. accurately predicting recidivism) and transparency, where the model is transparent about features of race, religion, income, housing and ethnicity that are fed into the algorithm and about how these features in the algorithm are weighted in order to produce an outcome about the risk of recidivism. Reflective equilibrium for overcoming these types of value conflict is a good starting point, but it is certainly limited. One problem is that it seems to assume that conflicts can indeed be always fully or partially overcome. However, this is not necessarily the case. While it can be useful to engage in reflective equilibrium exercises, we should equip students with other possibilities. One obvious suggestion can be to specify in more detail the types of relationship that there can be between values or principles. Sometimes principles can promote one another in a sort of feedback loop; other times they complement one another; but there might be cases where they stand in a tradeoff relation, and such tradeoffs can be unsolvable. Beth and Tom's module could generate a discussion on striking a balance between competing values, such as accuracy vs. efficiency. For example, this can be conceptualized as balancing between representational accuracy and reducing overall accidents. The module could be framed as dilemma in the design of the training scheme: deciding on whether to allocate training resources to identify darker skin pedestrians, so that they would not be disproportionately involved in accidents caused by autonomous vehicles; or whether to allocate the same training resources to reducing overall accidents, the result of which will maintain disproportionate harm to dark-skinned pedestrians.[3] More concretely, the module can explore the relationship between data on accidents and socio-economic characteristics. It would first ask students to design a data set for training an autonomous vehicle to avoid accidents. Then the students would be asked to find correlations and differences between a victim's ethnicity and various other variables in the data, and subsequently to write a short response to the question, "With respect to race, skin colour and other variables in the data, how could bias in the data or data collection be impacting or causing these differences?" The final part of the module would include an assignment of building three predictive models from the data that leave out race and other correlating variables in different ways in order to see what impact different variables are having on the model.

The challenges intrinsic to discretionary judgements of the sort are very instructive because they show that moral problems are not like engineering problems. While the latter has usually an accepted range of solutions, moral dilemmas might not. Still, addressing the moral tensions in a systematic way (e.g. efficiency vs. autonomy; fairness vs. accuracy; personalization vs. solidarity; convenience vs. dignity) and discussing potential avenues for addressing these tensions (Whittlestone et al. 2019) could provide a useful tool for engineering students. On a different vein, another strategy for dealing with value conflicts is to acknowledge the existence of 'irresolvable conflicts', and turn to social or political mechanisms such as compromise or deliberative democracy (Petersen 2021). This strategy belongs in the political sphere, to which we turn to now.

---

[3] We are grateful to Lotem Elber-Dorozko for pointing out this potential tradeoff.

## 4. The political dimension

We call the third perspective or dimension 'political', understood as the social and political values and institutions within which the moral dimension is realized. The political dimension is important when teaching applied ethics to engineers because as practitioners, engineers yield significant power in shaping, through the design of technology, social and political values and institutions. From this point of view, this dimension is about teaching the students to understand and appreciate their unique role in shaping the social and political structures of society through technology - to cultivate what we call a 'civic sensibility' by making students aware of how they might potentially contribute to undermining democracy if they participate in designing technology platforms that disseminate misinformation, or if they design a biased algorithm that reinforces structural injustice. But what is politics exactly?

Politics, in the normative sense, concerns what we all together as members of a community owe other members of the community, when we act in and on behalf of the artificial collective person which is the state (Dworkin 2011). This includes things like designing basic principles that will justify a particular form of state, formulating and protecting certain inalienable rights, deciding on how society's material resources should be shared among its members (Miller 1998).

The political dimension is indispensable, we argue, in the context of technology ethics pedagogy. For example, while many researchers and developers are beginning to take the problem of algorithmic bias seriously, the task of developing fair algorithms is still perceived as a technical task. Yet decisions on what would count as fair algorithms involve a choice between competing values. This is essentially a *political* question (Wong 2020), as it involves questions of authority, legitimacy, resolving disputes and power dynamics.

In the context of algorithmic wrongs, what matters is not primarily the identification and regulation of such wrongs, but more significantly "how algorithms are implicated in new regimes of verification, new forms of identifying a wrong or of truth telling in the world" (Amoore 2020, 5–6). As such, the political dimension is important because it shows how algorithmic operations and other technological developments in coding and machine learning may alter the basic terms and conditions under which moral norms are considered (Lewis 2022).

There are two dominant ways of explaining the distinction between morality and politics: the first explanation is the 'domain view'. This view holds that both morality and politics are concerned with the same normative questions, yet their focus differs: morality is concerned with individual conduct, whereas politics is concerned with institutional design and social organization. The second justification is the core-value view. On this view, morality is concerned with what is right, just and good, whereas politics is concerned with legitimacy, order, stability (List and Valentini 2020),[4] and power relations. Both these explanations of the distinction between morality and politics will serve us in this paper.

*4.1. The domain view: Basic structure of society and structural injustice*

The 'domain view' that normative politics (politics hereafter) is concerned with the institutions, social structures and social norms that govern morality at the level of the community is captured by the 'basic structure of society': "the interconnected system of rules and practices that define the political constitutions, legal procedures and the system of trials, the institution of property, the laws and conventions which regulate markets and economic productions and exchange, and the institution of the family" (Freeman 2003, 3). Realizing values such as justice and equality is primarily the purview of the basic structure, since it comprises a society's major social, political, and economic institutions. While individuals will be assigned a duty to support just institutions, within the framework established by those institutions, they will be able to lead their lives in such a way as to honour the values appropriate to small-scale interpersonal relationships (Scheffler 2005, 236). In other words, individuals are expected to uphold just institutions (the political dimension), in order that within those institutions they can choose how to live their lives (the ethical dimension). Values of justice and fairness are realized through the basic structure, since the basic structure should be comprised of just institutions, and upheld by just citizens.

Technology forms part of the basic structure, because it shapes many of the basic structure's components, such as markets, laws and social institutions such as the family. Furthermore, while the basic structure contains social and technical elements, at least arguably, these elements interact dynamically to constitute new forms of stable institutional practice and behavior (Gabriel 2022).

Furthermore, the domain conception of politics focuses on politics as a site of structural injustice: a systematic threat of domination or deprivation of the means to develop and exercise one's capacities, at the same time that these processes enable others to dominate or to have a wide range of opportunities for developing and exercising capacities available to them (Young 2011). Injustice, in this context, is interpreted as the terms of the power relations between agents in different positions created by social structures (Claassen and Herzog 2021). Structural injustices are not merely the result of an agent's ill will or personal failings. They are the interacting forces of conditions like poverty, precarious employment, bad housing and transportation policy, an inadequate welfare state, and various decisions by individuals, however defensible, constraining a person's options and leaving them unable to access basic goods (Sankaran 2021). In a state of non-domination, individuals have autonomous agency (Claassen and Herzog 2021), to lead their lives according to the ethical principles they adopt and pursue.

---

[4] For a discussion on possible distinctions between morality and politics see (List and Valentini 2020).

Importantly, in structural injustice individuals may be blameless, yet the schemes, in combination with norms and background conditions, systematically prevent some from developing their capacities (Young 2011). This feature of politics—the blamelessness of individual action combined with structurally unjust background conditions—is particularly relevant to discussions on the issue of individual responsibility for algorithmic outcomes. Structural bias entrenched by algorithms is another example of technology sustaining structural injustice.

The domain view of politics could feature in Beth and Tom's module by introducing the possibility of algorithmic structural discrimination in autonomous vehicle performance. Consider how the application of an algorithmic rule of behavior in a particular driving scenario, aggregated and accumulated across the autonomous vehicle system, could create patterned outcomes of convergence, possibly resulting in a disposition towards structurally bias, if the rule maps onto individual or group characteristics that are already discriminated against, or that discriminating against is morally impermissible (Liu 2018). The module could therefore work with the students on identifying the process whereby an algorithmic rule for the behaviour of an autonomous vehicle unit might generate system-wide unforeseen results that entrench structural injustice. This would also illustrate Young's point above, showing the students how individual actions may be blameless, yet the schemes, in combination with norms and background conditions, systematically discriminate against other individuals or groups.

Embedding the domain view of the political dimension in the engineering curricula can equip students and future professionals with the skills necessary for participating in society and upholding just institutions. In other words, embedding the political dimension in the engineering curricula is important for equipping students and future professionals with *civic sensibilities* (given the impossibility of full-blown virtues): to act in a specific context with a civic sensibility like civic attention is to act competently as a citizen or member of a political community (van den Brink 2013).

For the engineer, manifesting civic sensibilities has two senses, one as the engineer *qua* member of society, and the second as the engineer *qua* member of a professional community. On the one hand, the *engineer qua member of society* exemplifies a civic attention by upholding the just basic structure of society through toleration, respect for rights, respect for others' autonomy and for institutions that administer justice. Another manifestation of civic attention is the active participation in the political process of self-determination, sharing in public life (Hartley and Watson 2014). In the context of technology, this means that the engineer takes care not to participate in practices or belong to institutions that undermine the basic structure of society or that reinforce structural injustices. On the other hand, the engineer *qua* member of society can also take care to voice concerns to one's employer about technologies aimed at intrusive or predatory surveillance, or platforms that control the flow and dissemination of information in undemocratic ways. A civic sensibility like civic attention could also manifest in ensuring that the technology they develop enables *others* to participate fully in public life. For the engineer *qua* professional, civic virtue would manifest in things that are intrinsic to or constitutive of the profession. For example, training machine learning systems with attention to structural inequalities, or designing sorting algorithms with the background institutions in mind.

*4.2. The core-value view: legitimate authority*

On the core-value view, politics is an attempt to provide order via authority and legitimate coercion in conditions of disagreement. Whereas morality is concerned with the question of what we ought to do, this view of politics is concerned with the question of what we ought to do *when we disagree about what we ought to do* (Sleat 2016). Politics, therefore, must settle through authority and law issues that cannot be settled through reason or morality (Rossi and Sleat 2014), through the prism of legitimate authority and legitimate coercion.

Beth and Tom's ethics module could highlight the core-value view of the political dimension by addressing the autonomous vehicle 'fair distribution of risk' problem through Rawlsian 'Public Reason' (Binns 2018; Brändle and Schmidt 2021), a mechanism for deciding on justifiable principles (Rawls 1996). Recall that on the core value view, politics is the domain of what we do when we disagree, as well as legitimacy of authority. In this context, the module can help students reflect on two types of legitimacy:

1) the legitimacy of impacted stakeholders in participating in the design of the algorithm. For example, letting passengers of self-driving cars, or even pedestrians, set at least some of the driving parameters themselves as a way of achieving greater respect for reasonable pluralism, individual autonomy, and legitimacy (Himmelreich 2020).

2) the legitimacy of the students' authority as future engineers who create risk-distribution probabilities by designing algorithms. One can start by noticing how autonomous vehicles might radically change mobility, and given the magnitude of such changes, one may start asking on the ground of which authority engineers are entitled to implement such big changes. This can lead to a discussion on whether AI developers' authority needs to be ratified in a public reason mechanism, or some other political mechanism (e.g. expert panel; democratic vote).

## 5. Making the connections between the moral dimension and the political dimension in embedded ethics

In Sections 2, 3, and 4, we have developed a tripartite framework to design comprehensive ethics modules within engineering curricula. A summary of the framework can be visualized below, in Figure 1.

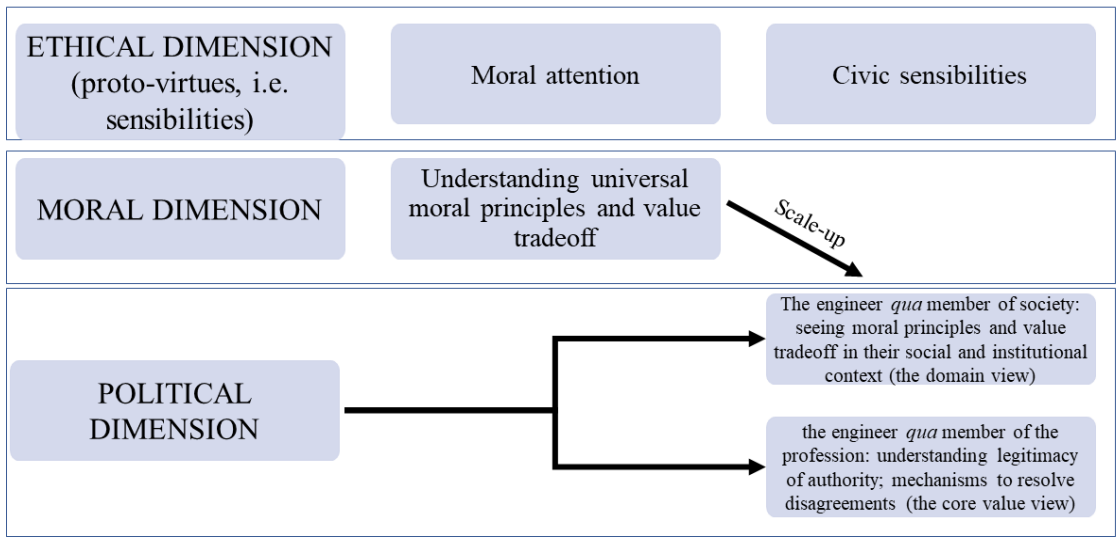

**Figure 1.** Representation of our tripartite framework for designing ethics modules

One advantage of our framework is that it neatly distinguishes between a 'moral' dimension and a 'political' dimension, which roughly corresponds respectively to the individual level and the collective level. These two dimensions, especially in contexts like AI, are often conflated or treated as equal. Normative concepts and values such as justice; fairness; responsibility; well-being; autonomy; privacy; democracy are often lumped together in the same analytic framework [see for example (Brey 2018; Fiesler, Garrett, and Beard 2020; Boldt and Orrù 2022; Himmelreich and Köhler 2022; Veluwenkamp et al. 2022), but contrast with (Rodríguez-Alcázar, Bermejo-Luque, and Molina-Pérez 2021)]. But putting into practice, it may be challenging to separate the moral and political dimensions in teaching the module. Explaining the distinction in depth can clarify this aspect.

      To illustrate the distinction between the moral and the political dimensions, consider the following artificial example: we imagine a module that aims to introduce the concept of unfairness in machine learning, by analyzing two distinct groups: the yellows and the greens. The module demonstrates the myriad ways in which an algorithm could generate unfair results favouring the yellows over the greens, and discusses the reasons why this unequal treatment would be morally wrong. This would be a moral dimension of bias and unfairness. A political dimension would discuss bias and unfairness through the context of the social and political structures that generated the conditions within which the greens are given lesser treatment compared to the yellows. Therefore, one can say that the moral dimension discusses why something is wrong, while the political dimension appreciates how something deemed morally wrong has ended up being structural, as well as contextualizing the wrongness as it manifests in institutions and political/social practices and norms. Another example of the importance of distinguishing between the moral and the political is the different conceptions of the concept 'responsibility' in AI. 'Responsibility' can denote a personal moral responsibility for a particular line of code, and it can also denote collective responsibility, as in the "problem of many hands", where responsibility lies at the hands of

the collective, but no individual can be held morally responsible (van de Poel, Royakkers, and Zwart 2015), especially when 'black box' AI is involved.

In the context of the autonomous vehicle module, Beth and Tom could make the connection between the moral and the political explicit by using the trolley problem as follows: for the moral dimension, the module could use trolley problems as a possible mechanism for reasoning about the justifications for moral decision-making. For the political dimension, the module could draw attention to the need to design the trolley problem so as to account for cumulative effects, which potentially lead to compound inequalities and discriminatory outcomes (Liu 2017). Thus the students will be encouraged to consider the societal context within which trolley problem scenarios are integrated into the autonomous vehicle's training scheme.

Designing modules that highlight both the moral and political dimensions can also sensitize the students to their civic duties as agents of responsible innovation. Beth and Tom's module could do this by asking the students to design a fair algorithm for risk distribution while reflecting on the methodology of obtaining the ethical input for the training scheme (using moral experts, or using observed behavior, or using stated preferences), and on the legitimacy of the agents that decide on ways to integrate ethical values into the training dataset.

However, a discussion on the legitimacy of decision-making agents, or on developers' authority in designing route-finding or risk-mitigation algorithms highlights a further distinction between the moral and the political dimensions. One of the challenges in creating a cohesive module is connecting each of the dimensions to the content of the course material, maintaining a connection between the technical skills and practical work that the students are undertaking, and the ethical, moral and political dimensions. The political dimension that highlights civic sensibilities (especially the responsibilities of the engineer as a member of society) and questions of legitimacy, is less directly connected to the technical material of the course, compared to the moral dimension. Let us illustrate this. In the autonomous vehicle module, the moral dimension can be taught using an assignment that directly deals with the design of fair algorithms. For example, one can assign a task such as "propose a principle of fairness for risk distribution that will be integrated in the algorithm, and justify this principle". The political dimension, on the other hand, has less to do with the design of the algorithm itself, and more to do with an opportunity to reflect on the civic role of the students as future practitioners that will be implementing large-scale changes in society. Here, the issue that the module can encourage the students to grapple with is the power dynamics between the individuals who design and implement algorithms, the persons or groups that have to live with the implications of these technologies (Ferreira and Vardi 2021; Herzog 2021; O'Neil 2016), and the collective decision-making procedures that are necessary for settling such issues (Zandvoort 2005).

But this is not to say that it is not possible to make an explicit connection between the technical aspects of the course material and the political dimension. For example, one can highlight the possibility that algorithm design is in a position to repair structural injustice. Let's say that the module uses Rawls' 'Original Position' – a methodology of reasoning about principles of justice from a position of ignorance regarding one's own status in society (Rawls 1971) as a thought experiment. Implementing the original position in the design of algorithms entails that the algorithms ought to be designed not only to remove bias or discrimination, but to actually improve the situation of the worst off in society. As such, by implementing the original position in the design of algorithms, the students cultivate civic responsibility by considering their roles in fighting structural injustices through the design and implementation of algorithms.

Finally, the political dimension helps to expand the horizons beyond the 'preventive ethics' state of mind. Currently, many technological developments are targeted primarily to individual issues rather than social ones. For example, while promoting human health through genomic technologies is a worthy goal, it frames health and disease as primarily an individual issue rather than a social issue (de Melo-Martín 2022), thereby blinding society to the moral duty that we have to come up with solutions at the societal level (whether technological or otherwise). The political dimension can also help dispel the existing ideal, prevalent in engineering ethics, of the "heroic engineer": someone who is both quite individualistic and at the same time strong enough to deal with whatever moral challenge they may face (Basart and Serra 2013). Teaching engineers to consider the political dimension to expand their moral sensibility beyond the individual-focused approach, could be done by introducing them to considerations beyond individual actions. For example, introducing the idea of the 'tragedy of the epistemic commons' in the context of fake news and echo chambers on social media (Rini 2017). The epistemic commons is "the stock of facts, ideas, and perspectives that are alive in society's discourse"(Joshi 2021, 8), yet it is arguably more than the facts, ideas and perspectives, as it includes the sharing of these facts, ideas and perspectives. This sharing attribute of the epistemic commons can be characterized by what Charles Taylor (1995) identifies as having an 'irreducibly social' component. Since fake news and echo chambers threaten the epistemic commons (Rini 2017; Nguyen 2020), introducing the students to the irreducibly social character of knowledge distribution on social media can help sensitize them to the political dimension of the problem. A concrete assignment in a module on social media and the epistemic commons could ask the students to design a social media platform that includes mechanisms for immediate detection of fake news, or that deliberately exposes users to a diverse 'diet' of reliable sources of information. By drawing attention to the collective aspects of technological innovation, the political dimension can help students critically examine whether the problem they are learning to solve lies in the individual realm or in the social realm, and tailor the solution accordingly.

## 6. Reflections on future research development

Embedded ethics is an emerging pedagogical approach, and as such there are many open questions regarding its practice and efficacy. Further research is necessary to explore the relationships between the implementation, teaching, assessment methods, the goals and theoretical frameworks applied in computer science and engineering ethics education (Keefer et al. 2014; Martin, Conlon, and Bowe 2021; for some recent attempts see (Kopec et al. 2023; Horton et al. 2022; 2023). Applying these questions to the pedagogy of embedded ethics requires a separate analysis. Here we briefly mention two issues that would benefit from further investigation, that are directly related to the focus of this paper, namely the need to distinguish between the ethical, moral and political dimensions.

*6.1 Domain expertise*

The hypothetical module we have used as illustration in this paper is co-taught by a philosopher and an engineer. While this is one model for creating an embedded ethics curriculum, it is not the only one. Embedded ethics is inherently multidisciplinary (Grosz et al. 2019), therefore it is necessary that the module is developed by relevant experts who each has their own domain expertise—in engineering or in philosophy. However, the module itself could be taught by a philosopher, (as is the practice in the Embedded EthiCS pioneering program, Grosz et al. 2019), or can be taught by the engineer, once they feel they have mastered the necessary ethical skills for an effective delivery. Each model has advantages and drawbacks. The first option (i.e., a module taught by the philosopher) can be perceived as following medical ethics, where an ethics specialist provides the domain expertise, but can be criticized on the grounds that it frames the ethical content as outside the responsibility of the engineer. The second option (i.e., a module taught by the engineer) can be perceived as aspiring to an ideal of a communal mode of shared ethical responsibility, where the ethical wisdom is distributed across agents (makers, developers, CEOs, regulators), and not specialized (Howard 2020). This approach sends the message that ethical issues are a fundamental and integral part of computing, and not a tangential topic that they can take a course on elsewhere (Pasricha 2023). Interestingly, the latter approach is rooted in the commitment to the cultivation of civic virtues in engineers (Howard 2018), which would be aligned with the integration of the political dimension into the module content. Nevertheless, training engineering instructors to teach ethical content might prove challenging as they would need to gain a certain degree of domain expertise in philosophy and ethics. Ultimately, determining which model would be more effective in the cultivation of the ethical, moral and civic sensibilities should probably be empirically tested.

*6.2 Integrating the ethical content within the technical material*

The location of the module within the structure of the whole course probably matters for the effective integration of ethical reasoning skills. For example, a module on integrating moral considerations into autonomous vehicle reward functions requires that the students are familiar with the technical aspects of reward functions before they can consider the moral considerations as part of the machine learning's training process. Alternatively, the module could be located closer to the latter end of the course after several of the technical skills have been addressed. As mentioned earlier, the political dimension might be more challenging to integrate into the technical material, and this could affect the decision on where to locate the module, so that it achieves its learning goals. Again, this too is an issue that could benefit from empirical investigation.

*6.3 Other challenges*

Our tripartite framework has been illustrated mostly through examples related to AI tools. This idiosyncrasy is a consequence of the area of specialization and background of the authors of this article. But there is nothing in our tripartite framework which is intrinsically exclusive of AI tools. In fact, we hope that our framework will be useful to ethicists and engineers working in other disciplines.

Another challenge is related to measuring learning outcomes. Our framework is agnostic to this problem, in the sense that it does not prescribe any specific method to assess students' learning, at least not explicitly. As such, it can accommodate different methods, but which one is the most suitable, is a topic of future research.

**7. Conclusions**

Calls to integrate ethics into technology, especially when they come from more technical contexts, tend to have confusing learning goals (Johnson 2017; Bezuidenhout and Ratti 2021) and conflate political and moral concerns under the same banner (Brey 2018; Fiesler, Garrett, and Beard 2020; Boldt and Orrù 2022; Himmelreich and Köhler 2022; Veluwenkamp et al. 2022). The lack of clearly specified goals jeopardizes the effectiveness of these calls for integration, and the conflation of different dimensions is also problematic, given that they bear upon different questions that should be kept distinct.

Our framework has the advantage of addressing both issues. Given that moral and political questions are distinct questions, then the technical answers would differ accordingly. Furthermore, the moral and the political emphasize the cultivation of closely related yet distinct sensibilities (e.g. moral attention, civic attention). Therefore, keeping the two dimensions distinct, while at the same time using both to complement one another, is fundamental, in order to address the agency of engineers and computer scientists and their capacity to deliver just and beneficial solutions. This can be enabled at the individual level, applying the ethical and moral dimensions, and at the collective level, applying the political dimensions, for example by getting engineering and computer science students to "second guess", or reconsider, the foundations of their fields (Lynch 2015).

We draw on Conlon (2022) to suggest two ways that embedded ethics could help computer science and engineering students reconsider the foundations of their field: first, in order for the political dimension to be contextualized properly, it is necessary for ethics modules to be explicitly connected to a boarder curriculum that exposes the structural dimensions of social and political life (e.g. in 'stand alone' courses). Second, having modules that focus on the political dimension, casting the normative lens on the question "what needs to change?" rather than on "what would you do"? In this way, the module can help students grapple with the agency of engineers and computer scientists that can be enabled at the political level, and their responsibility to participate in shaping just structures, including, importantly, changing employment relationships and harmful hierarchical structures in corporate entities within which most engineers and computer scientists work.

A second advantage of our framework is that it demonstrates the importance of the distinction within political philosophy regarding what the political is. Recall that the domain view of politics is concerned with the institutions, social structures and social norms that govern morality at the level of the community. On this view, moral questions answered at the institutional level are carried over from the individual level. The core value view, on the other hand, treats the moral and the political as distinct sets of questions. While at the moral level ethics will be concerned with what the individual ought to do with respect to technological developments, the political level is a site for reflection on resolving disagreements about these such technological developments. As such, the analytical distinction between the moral dimension and the core value view of the political dimension is much more pronounced. The practice of integrating ethics into the computer science and engineering curricula itself has a political dimension, in the sense that it is not restricted to the attitudes or responsibilities of individual instructors, but is affected by the institutional measures and policies set by the university, as well as the cultural milieu in which they ethical content is being taught, shaping collective identities about what is valued in engineering education (Martin, Conlon, and Bowe 2021). Making the political and moral dimensions explicit in embedded ethics curricula might therefore help make visible the political dimension in the practice of embedded ethics as a pedagogical approach. This is useful for both students and instructors.

Finally, our framework provides ethics instructors with clarity regarding the question they are addressing, and help them maintain a balance between the moral and the political. While it is impossible to capture all the moral and political implications of a technology in one module, knowing which dimension one is addressing in that module can help build a comprehensive embedded ethics *program*, such that ensures that students are exposed to both moral and political dimensions across the curriculum. In this article, we have proposed a framework to overcome these difficulties and to properly integrate ethics modules in technical curricula.